# Relation between the weak itinerant magnetism in $A_2$Ni$_7$ compounds ($A$ = Y, La) and their stacked crystal structures

J.-C. Crivello and V. Paul-Boncour*

CNRS, UPEC, ICMPE (UMR7182), F-94320 Thiais, France

*E-mail: paulbon@icmpe.cnrs.fr

**Abstract**

The weak itinerant magnetic properties of $A_2$Ni$_7$ compounds with $A$ = {Y, La} have been investigated using electronic band structure calculations in the relation with their polymorphic crystal structures. These compounds crystallizes in two structures resulting from the stacking of two and three blocks of [$A_2$Ni$_4$ + 2 $A$Ni$_5$] units for hexagonal 2$H$-La$_2$Ni$_7$ (Ce$_2$Ni$_7$ type) and rhombohedral 3$R$-Y$_2$Ni$_7$ (Gd$_2$Co$_7$ type) respectively. Experimentally, 2$H$-La$_2$Ni$_7$ is a weak itinerant antiferromagnet whereas 3$R$-Y$_2$Ni$_7$ is a weak itinerant ferromagnet. From the present first principles calculation within non-spin polarized state, both compounds present an electronic density of state with a sharp and narrow peak centered at the Fermi level corresponding to flat bands from 3$d$-Ni. This induces a magnetic instability and both compounds are more stable in a ferromagnetic (FM) order compared to a paramagnetic state ($\Delta E \approx$ -35 meV/f.u.). The magnetic moment of each of the five Ni sites varies with their positions relative to the [$A_2$Ni$_4$] and [$A$Ni$_5$] units: they are minimum in the [$A_2$Ni$_4$] unit and maximum at the interface between two [$A$Ni$_5$] units. For 2$H$-La$_2$Ni$_7$, an antiferromagnetic (AFM) structure has been proposed and found with an energy comparable to that of the FM state. This AFM structure is described by two FM unit blocks of opposite Ni spin sign separated by a non-magnetic layer at $z$ = 0 and ½. The Ni (2$a$) atoms belonging to this intermediate layer are located in the [La$_2$Ni$_4$] unit and are at a center of symmetry of the hexagonal cell ($P6_3/mmc$) where the resultant molecular field is cancelled. Further non-collinear spin calculations have been performed to determine the Ni moment orientations which are found preferentially parallel to the **c** axis for both FM and AFM structures.

# 1. Introduction

Weak itinerant ferromagnets (wFM) and antiferromagnets (wAFM) are not very common and have raised a large and ongoing interest as they are near the onset of magnetism, present large spin-fluctuations and are often close to quantum critical point and even superconductivity [1-3]. Among the wFM, several studies have been performed on $ZrZn_2$ [4, 5], $Ni_3Al$ [6-10], $Fe_2N$ [11, 12] and $AsNCr_3$ [13]. wAFM are even less common than wFM and have been observed in some compounds like $TiBe_2$ [14-16], TiAu [17-20] and UN [21]. Weak itinerant magnets are generally characterized by high transition temperatures and low magnetic moments related to their particular electronic properties.

Recently Singh [22] has compared wFM classical faced-centered-cubic Ni substituted by Al forming the $Ni_3Al$ (75 % Ni) with wFM $Y_2Ni_7$ which is also a Ni rich compound (78 % Ni). $Y_2Ni_7$ is an interesting itinerant magnet which has deserved both detailed theoretical and experimental studies [23-25].

With the same stoichiometry than $Y_2Ni_7$, $La_2Ni_7$ has been the subject of many experimental studies revealing a weak itinerant antiferromagnetic ground state, whose origin has not been clearly solved until now. Due to the very few number of wAFM compounds, it appears of particular interest to describe the AFM structure of $La_2Ni_7$ and to understand the origin of its specific behavior.

$La_2Ni_7$ crystallizes preferentially in the hexagonal $Ce_2Ni_7$ structure type and $Y_2Ni_7$ in the rhombohedral $Gd_2Co_7$ structure type. Hexagonal $La_2Ni_7$ is a wAFM with $T_N = 50$ K and undergoes a metamagnetic transition toward a wFM state [26-30], whereas rhombohedral $Y_2Ni_7$ is a wFM with $T_C = 53$ K [22, 23, 25, 31, 32]. Both compounds have in common low value of the mean Ni moments (0.06 to 0.11 $\mu_B$/Ni) [25, 30] and relatively large ordering temperatures, characteristic of the weak itinerant magnetism. They have also effective Ni moments around $\mu_{eff} = 0.8\sim1$ $\mu_B$ and positive paramagnetic Curie temperatures $\theta_p$ between 50 and 70 K derived from the analysis of their magnetic susceptibility in the paramagnetic range [23, 25, 27, 30]. This is rather surprising for $La_2Ni_7$, as a negative $\theta_p$ is expected for an antiferromagnetic compound.

Several experimental studies have confirmed the wAFM behavior of $La_2Ni_7$ at low field, but the attempts to perform neutron powder diffraction (NPD) experiments at low temperature [33] did not allow to determine its microscopic magnetic structure. The absence of magnetic peaks in the neutron patterns below $T_N$ was attributed to the itinerant character of the antiferromagnetism and the low values of the Ni moments.

In order to solve the origin of the weak antiferromagnetic ground state of $La_2Ni_7$ as well as its metamagnetic transition toward a ferromagnetic structure, we have performed first principle calculations using collinear and non-collinear spin polarized Density-Functional Theory (DFT). Taking into account the particular geometry of the $La_2Ni_7$ crystal structure, we will propose an antiferromagnetic structure which can explain, for the first time, many of the magnetic experimental particularities of this compound.

The magnetic and electronic properties of $La_2Ni_7$ will be compared with those of $Y_2Ni_7$ all along this paper. A more general discussion on the specific itinerant magnetic behavior of $La_2Ni_7$ compared to other intermetallic compounds with large spin fluctuations will be introduced.

## 2. Structural description and methodology

$La_2Ni_7$ and $Y_2Ni_7$ crystallize in a $Ce_2Ni_7$ type hexagonal structure ($P6_3/mmc$ space group) and $Gd_2Co_7$-type rhombohedral structure (trigonal symmetry, $R\text{-}3m$ space group) respectively as presented in figure 1. These phases result from the stacking of $[A_2B_4]$ and $[AB_5]$ units according to the rule $[A_2B_4 + n.AB_5]$, where $n$ is an integer. If $n = 2$, the stacking leads to the $A_2B_7$ formation in two polymorphic forms: the hexagonal $Ce_2Ni_7$ cell [$2H$] contains two blocks of $[A_2B_4 + 2\ AB_5]$, whereas the rhombohedral $Gd_2Co_7$ cell [$3R$] contains three blocks of $[A_2B_4 + 2\ AB_5]$ in the hexagonal description as shown in figure 1 [34]. For the sake of simplicity, the $[A_2B_4]$ subunit will be renamed as $[AB_2]$ in the following. Both hexagonal and rhombohedral structures are strongly anisotropic, with large values of the $c$ parameter compared to the $a$ parameter ($a = 5.062$ Å, $c = 24.71$ Å for $2H$-$La_2Ni_7$; $a = 4.946$ Å and $c = 36.26$ Å for $3R$-$Y_2Ni_7$). Both cell contains 2 $A = \{Y, La\}$ and 5 $B = $ Ni different Wyckoff sites (table 1) which are characterized by their positions along the $c$ axis forming $A$ and Ni layers. Each $A$ site is located either in one $[AB_5]$ or one $[AB_2]$ units, whereas each Ni site is located either in one of these units or at the interface between 2 units. Regarding the $Ce_2Ni_7$ type structure, it displays a mirror inversion of the $[AB_2]$ units at $z = 0$ and $z = 1/2$.

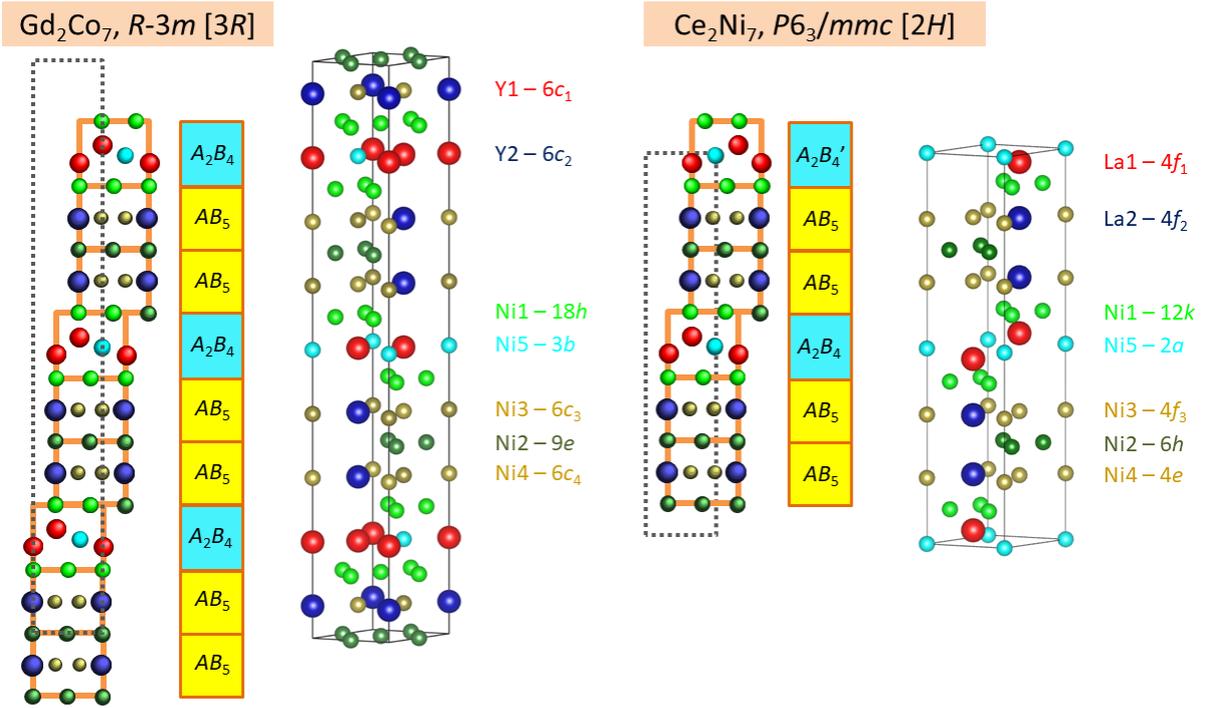

Figure 1: Comparison of the rhombohedral and hexagonal structures of $A_2Ni_7$ compounds showing the stacking of $[AB_2]$ and $[AB_5]$ units and the $A$ and Ni atom positions.

The electronic structure was calculated for the ordered $La_2Ni_7$ and $Y_2Ni_7$ compounds in both hexagonal and rhombohedral structures. In the frame of the DFT, the pseudo-potential approach using the VASP package [35, 36] was considered using projector-augmented wave [37] with a 600 eV cut-off energy and a high $k$-mesh density (at least 250 k-point in the irreducible Brillouin zone). Several exchange and correlation (XC) functionals have been considered, such as local density approximation (LDA) with parametrization from Perdew and Zunger [38], the generalized gradient approximation with the PBE functional [39, 40] and a meta-GGA functional, the strongly constrained and appropriately normed semilocal density functional (SCAN) [41], including the second derivative of the electron density. In addition, an empirical coulomb interaction (Hubbard) was introduced ($U = 5$ eV, $J = 0$) to test the influence of GGA+U calculation [42]. Preserving the original crystal symmetry, each structure has been fully relaxed within several magnetic ordering such as ferromagnetic and antiferromagnetic structure using the electronic collinear and non-collinear spin-polarization. For the latter, the most stable spin moments orientation has been investigated in GGA-PBE (several moment directions have been tested such as parallel and perpendicular to $c$ axis). Final static calculation was done using the linear tetrahedron method with Blöchl corrections on relaxed structures [43]. The charge distribution on the atoms was investigated using Bader topological analysis within the "Bader" code developed by Henkelman et al. [44, 45].

## 3. Results and discussion

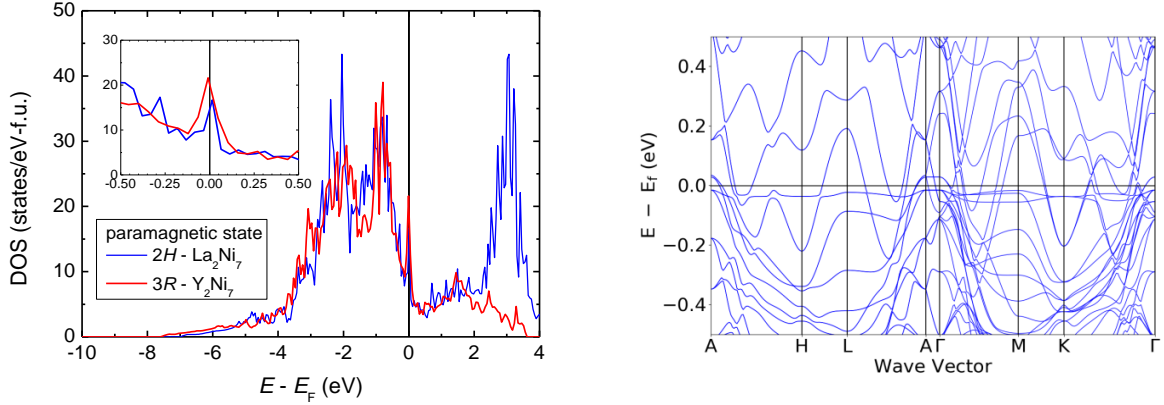

Figure 2 : Non-spin polarized density of state of 3$R$-Y$_2$Ni$_7$ and 2$H$-La$_2$Ni$_7$ with a zoom near E$_F$ in the inset (left) and corresponding electronic band structure of 2$H$-La$_2$Ni$_7$ zoomed near E$_F$ (right) in GGA-PBE. The full scale of the latter is given in supplementary materials.

The non-spin polarized (NSP) density of state (DOS) of hexagonal La$_2$Ni$_7$ (2$H$-La$_2$Ni$_7$) and rhombohedral Y$_2$Ni$_7$ (3$R$-Y$_2$Ni$_7$) are compared in figure 2 (left). The electronic DOS results in a large main structure dominated by the 3$d$ states from Ni. A charge transfer from the $A$ element to Ni is observed (1.2 e- per $A$ atom according to the Bader method), but is not sufficient to completely fill the Ni-bands and both compounds present a DOS with a sharp and narrow peak centered at the Fermi level ($E_F$). This result is in agreement with previous studies [22, 46, 47] and was also observed in wAFM orthorhombic TiAu [18, 19] and H$_3$S superconductor [48]. This peak corresponds to the flat $d$-bands in the electronic band structure as shown in figure 2 (right, zoom scale) and figure S1 (full scale) for 2$H$-La$_2$Ni$_7$. Such flat bands were also visible for the paramagnetic calculation of 3$R$-Y$_2$Ni$_7$ in ref. [22]. It can be noticed that the peak maximum is located at $E_F$ for Y$_2$Ni$_7$, while it is slightly shifted to higher energy for La$_2$Ni$_7$. The main contribution to this peak is thus due to the 3$d$ states of Ni and leads to a high DOS at $E_F$; $N(E_F)$ = 14.6 states/eV per f.u. for 2$H$-La$_2$Ni$_7$, smaller than that calculated for 3$R$-Y$_2$Ni$_7$ : $N(E_F)$ = 21.7 states/eV per f.u. (close to the 24 states/eV per f.u. from ref. [22]). Both values are large enough to follow the Stoner criterion [49] ($N(E_F)*I$ > 1 for $I \approx$ 0.8 eV/at usually taken for transition metals) and induce a lowering of the total energy by a shift of the minority and majority bands.

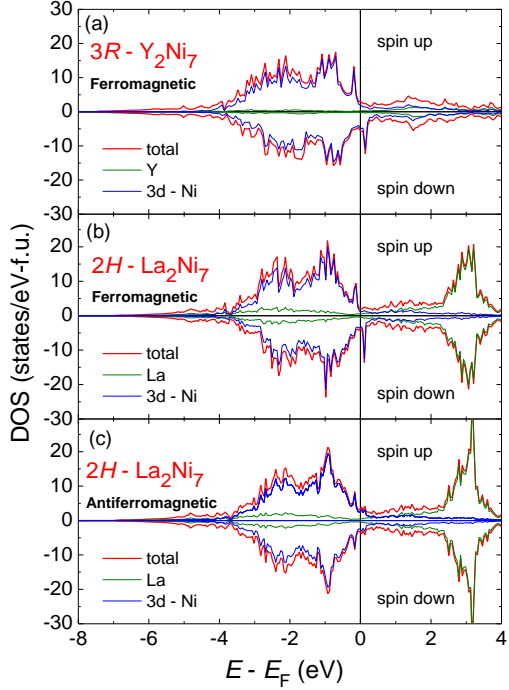

Figure 3: Calculated spin polarized DOS with partial 3d-Ni contribution for (a) ferromagnetic 3R-Y$_2$Ni$_7$ ($M$ = 1.24 µ$_B$/f.u.), (b) ferromagnetic 2H-La$_2$Ni$_7$ ($M$ = 1.08 µ$_B$/f.u.) and (c) antiferromagnetic 2H-La$_2$Ni$_7$ ($M$ = 0 µ$_B$/f.u.).

The calculated spin polarized DOS of 3R-Y$_2$Ni$_7$ and 2H-La$_2$Ni$_7$, assuming a ferromagnetic ground state, are displayed in figure 3a and figure 3b respectively. The calculated moments for each atomic site as well as the magnetic energies ($\Delta E$) are reported in table 1. The high density peak position is shifted to lower and higher energy for spin up and spin down bands respectively and $N(E_F)$ decreases due to its position located in a valley of the density of states. As a consequence the total energy of ferromagnetic state is lowered by 5 meV/ Ni atom compared to NSP calculated value for both compounds. This magnetic energy corresponds to a temperature of 58 K, which is close to the experimental ordering temperature of Y$_2$Ni$_7$ ($T_C$=53 K) [25].

The calculated moment for each atomic site in ferromagnetic state are reported in Table 1. The magnetic moments in Y$_2$Ni$_7$ are in good agreement with those calculated by Singh using a linear augmented plane-wave method [22]. It is noticeable, that the total calculated moments of 1.08 µ$_B$/f.u. for La$_2$Ni$_7$ and 1.24 µ$_B$ for Y$_2$Ni$_7$ are larger than experimental values (0.77 µ$_B$/f.u. and 0.43 µ$_B$/f.u. measured at 4.2 K and extrapolated from high field for $A$ = La and Y respectively) [25, 27]. This overestimation is due to the limit of our DFT calculations, which do not take into account the large spin fluctuations present in such compounds. The Ni moments contribute mainly to the total magnetization, whereas the induced *A* moments remain weak and are, as expected, of opposite sign compared to Ni moments.

**Table 1**: Calculated magnetic moments in hexagonal La$_2$Ni$_7$ (FM, AFM) and rhombohedral Y$_2$Ni$_7$ (FM). $\Delta E = E-E_0$ (NSP).

| 2*H*-La$_2$Ni$_7$ | | | | FM | AFM |
|---|---|---|---|---|---|
| Atom | site | Unit | neighbors | *m* (μ$_B$/at.) | *m* (μ$_B$/at.) |
| La1 | 4*f$_1$* | [*AB$_2$*] | 4 *A*+12 Ni | -0.026 | ±0.015 |
| La2 | 4*f$_2$* | [*AB$_5$*] | 2 *A*+18 Ni | -0.041 | ±0.041 |
| Ni1 | 12*k* | [*AB$_2$/AB$_5$*] | 5 *A*+7 Ni | 0.102 | ±0.095 |
| Ni2 | 6*h* | [*AB$_5$*] | 4 *A*+8 Ni | 0.287 | ±0.293 |
| Ni3 | 4*f$_3$* | [*AB$_5$*] | 3 *A*+9 Ni | 0.198 | ±0.186 |
| Ni4 | 4*e* | [*AB$_5$*] | 3 *A*+9 Ni | 0.197 | ±0.197 |
| Ni5 | 2*a* | [*AB$_2$*] | 6 *A*+6 Ni | 0.023 | 0 |
| $M_{total}$ (μ$_B$/f.u.) | | | | 1.08 | 0 |
| $\Delta E$ (eV/f.u.) | | | | -0.035 | -0.033 |
| 3*R*-Y$_2$Ni$_7$ | | | | FM | |
| Atom | site | Unit | neighbors | *m* (μ$_B$/at.) | |
| Y1 | 6*c$_1$* | [*AB$_5$*] | 2 *A*+18 Ni | -0.043 | |
| Y2 | 6*c$_2$* | [*AB$_2$*] | 4 *A*+12 Ni | -0.034 | |
| Ni1 | 18*h* | [*AB$_2$/AB$_5$*] | 5 *A*+7 Ni | 0.153 | |
| Ni2 | 9*e* | [*AB$_5$*] | 4 *A*+8 Ni | 0.315 | |
| Ni3 | 6*c$_3$* | [*AB$_5$*] | 3 *A*+9 Ni | 0.208 | |
| Ni4 | 6*c$_4$* | [*AB$_5$*] | 3 *A*+9 Ni | 0.216 | |
| Ni5 | 3*b* | [*AB$_2$*] | 6 *A*+6 Ni | 0.071 | |
| $M_{total}$ (μ$_B$/f.u.) | | | | 1.238 | |
| $\Delta E$ (eV/ f.u.) | | | | -0.035 | |

Further non-collinear spin calculations have been performed to determine the preferential orientation of the Ni moments in the FM structure. The Ni moments are stabilized in the direction parallel to the *c* axis: the energy difference between moments parallel and perpendicular to the *c* axis is about -30 meV/f.u..

The coordination polyhedra have been represented with their calculated magnetic vectors for the Ni1 to Ni4 sites in figure 4 and for the Ni5 site in figure 5 (right) for hexagonal La$_2$Ni$_7$ in the ferromagnetic state. In a first approximation, the magnitude of Ni moments should depend on the number of Ni neighbors and of the site symmetry. Assuming that the Ni moments should be larger for atoms surrounded by a larger number of Ni neighbors, the Ni3 and Ni4 atoms, which have similar coordination environment with 9 Ni neighbors and 3 *A* neighbors, are expected to have larger magnetic moment than the Ni2 atom surrounded by only 8 Ni neighbors, but the calculation indicates that the Ni2 moment is the largest (table 1). In fact, a closer analysis of the Ni–Ni pair distances shows that each Ni3 (Ni4) atom is surrounded by three Ni2 at 2.45 ± 0.5 Å, three Ni1 at 2.52 ± 0.5 Å and three Ni4 (Ni3) atoms distant of 2.92 ± 0.5 Å (at the same distance than the 3 *A* neighbors). The larger Ni4–Ni3 distance corresponds to weaker Ni–Ni bonds and smaller exchange interaction with Ni3 and Ni4 moments. On the other hand the Ni2 atoms are surrounded by 8 Ni at distances between 2.45 and 2.62 Å forming a more regular polyhedron with 4 Ni2 neighbors in the basal plane and 4 Ni3 (Ni4) neighbors in the equatorial plane. The Ni1 atoms are surrounded by 4 Ni1 in the basal plane, 1 Ni5, 1 Ni3 and 1 Ni4 with distances ranging between 2.45 Å and 2.62 Å. The Ni5 atoms are surrounded by 6 Ni1 atoms at 2.55 Å (figure 5 right).

In a simple molecular field approximation, the Ni–Ni exchange interactions can be described by an internal field $H_{exch}$ acting on the Ni site and resulting from the closest Ni neighbor moments (the influence of the Ni–La interactions can be neglected as they are much smaller):

$$H_{exch} = \lambda_{Ni-Ni} \cdot m_{Tot} \qquad (1)$$

where $\lambda_{Ni-Ni}$ is the Ni-Ni molecular field coefficient and $m_{tot} = \sum_j z_j \cdot m_j$ is the sum of the $z_j$ Ni neighbor moments $m_j$. Assuming that the Ni–Ni short range interaction is close for all Ni atoms and that the exchange integral are not very different for the Ni–Ni distances between 2.45 and 2.62 Å, it is possible to verify if there is any proportionality between the Ni moments on each site and the sum of its first neighbor moments. The variation of each Ni moment ($m_i$) versus $m_{Tot}$ shows two different behaviors for the Ni atoms related to the [*AB$_2$*] units and those belonging to the [*AB$_5$*] units (figure 6a) with a d$m_i$/d$m_{Tot}$ slope of 0.58 and 0.125 respectively. To explain this difference of behavior, which cannot be only due to the variation of Ni–Ni distances as they are comparable for Ni1 and Ni2 atoms and their next Ni neighbors, one should also consider the influence of the first La neighbors (2.93 Å ≤ $d_{Ni-La}$ ≤ 3.33 Å).

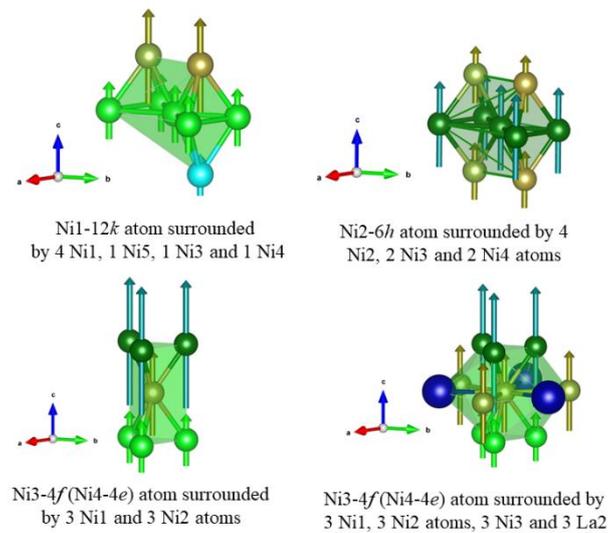

Figure 4: Coordination polyhedra around the Ni1, Ni2 and Ni3 (Ni4) atoms. For the Ni3 (Ni4) atoms two types of polyhedra have been represented for interatomic distances below 2.65 Å (left) and below 2.95 Å (right) The magnetic vectors have been represented with length proportional to their magnetic moments.

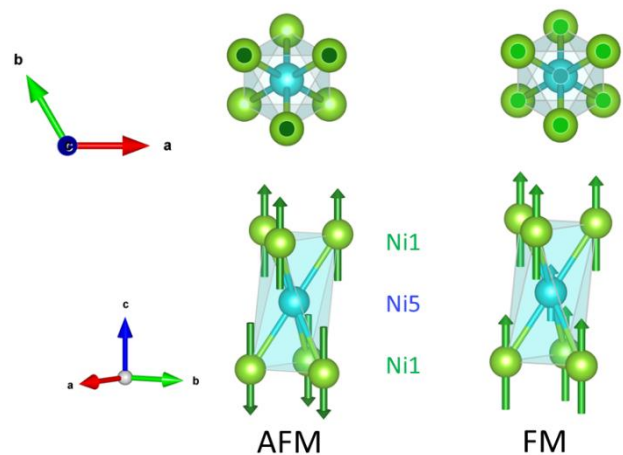

Figure 5: Coordination polyhedra of Ni5 atoms surrounded by Ni1 atoms with magnetic vectors in AFM and FM ordering perpendicular (top) and parallel (bottom) to the **c** axis.

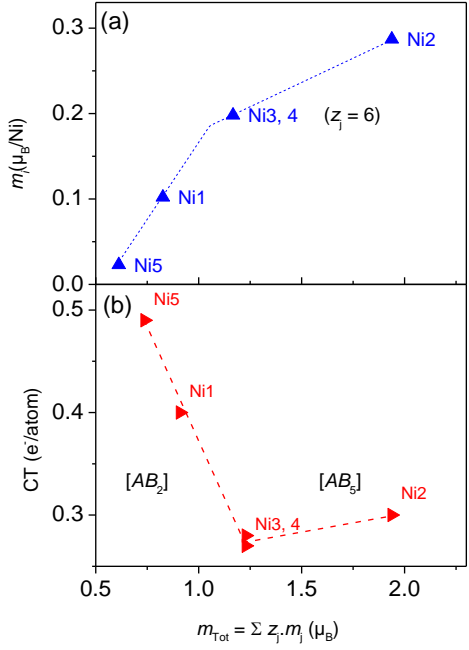

Figure 6: Ni moment and charge transfer CT from La to Ni versus the sum of the Ni moments in the first coordination sphere $m_{Tot}$ for each Ni site in $2H$-$La_2Ni_7$.

The electronic charge transfer CT from La to Ni atoms is calculated using the Bader method and is given as a function of $m_{Tot}$ in figure 6b. Similar to the $m_i$ individual moments, the CT presents two different behaviors for Ni in $[AB_2]$ and $[AB_5]$ units: a sharp decrease from Ni5 to Ni3,4 and an almost constant CT value for Ni3,4 and Ni2 atoms. The CT values are related to the number of La first neighbors (table 1). For Ni atoms with the largest CT values (Ni5 and Ni1) the additional electrons transferred from La atoms contribute to a progressive filling of the $d$-Ni bands. It yields a reduction of their local magnetic moment compared to the values extrapolated from Ni3,4 or Ni2 moment versus $m_{Tot}$.

As a consequence, the magnitude of the Ni moments can be related to their positions relative to the $[AB_2]$ and $[AB_5]$ units. For both structure types, it appears that the Ni moment is maximum for the Ni2 atoms located between two $[AB_5]$ units and minimum for the Ni5 ($2a$) atoms belonging to the $[AB_2]$ units, at the $z = 1/2$ plane containing the center of symmetry of the cell. This is illustrated in figure 7, where the hexagonal structure of $La_2Ni_7$ and the corresponding magnitude of the Ni moments along the $c$ axis are presented. Similar Ni moment distribution can be obtained for $3R$-$Y_2Ni_7$. It is also noticeable that the distribution of Ni magnetic moments along half the hexagonal cell forms a triangle. This is repeated twice, because the stacking $[A_2B_4$

+ 2 $AB_5$] is repeated two times in this hexagonal lattice, whereas the same sequence appears three times in 3$R$-Y$_2$Ni$_7$. This emphasizes a kind of modulation of the Ni moments along the $c$-axis.

**Table 2:** Heat of formation of $A_2$Ni$_7$ compounds in both allotropic hexagonal and rhombohedral structures in different magnetic states. For hexagonal La$_2$Ni$_7$ the calculations were done with different functionals and method as described in the text.

| Compound | XC- method | Structure | Magnetic state | $\Delta E = E(i) - E(NSP)$ (meV/f.u.) |
|---|---|---|---|---|
| Y$_2$Ni$_7$ | GGA | rhombohedral | NSP | 0 |
| | | | FM | -35 |
| | | hexagonal | NSP | 0 |
| | | | FM | -35 |
| | | | AFM | -33 |
| La$_2$Ni$_7$ | GGA | rhombohedral | NSP | 0 |
| | | | FM | -31 |
| | | hexagonal | NSP | 0 |
| | | | FM | -35 |
| | | | AFM | -33 |
| | non-colinear | $M$ parallel to z | FM | -34 |
| | | | AFM | -33 |
| | non-colinear | $M$ parallel to x, y | FM | -3 |
| | LDA | hexagonal | NSP | 0 |
| | | | FM | -22 |
| | | | AFM | -20 |
| | SCAN | hexagonal | NSP | 0 |
| | | | FM | -75 |
| | | | AFM | -72 |
| | GGA + U | hexagonal | NSP | 0 |
| | | | FM | -35 |
| | | | AFM | -32 |

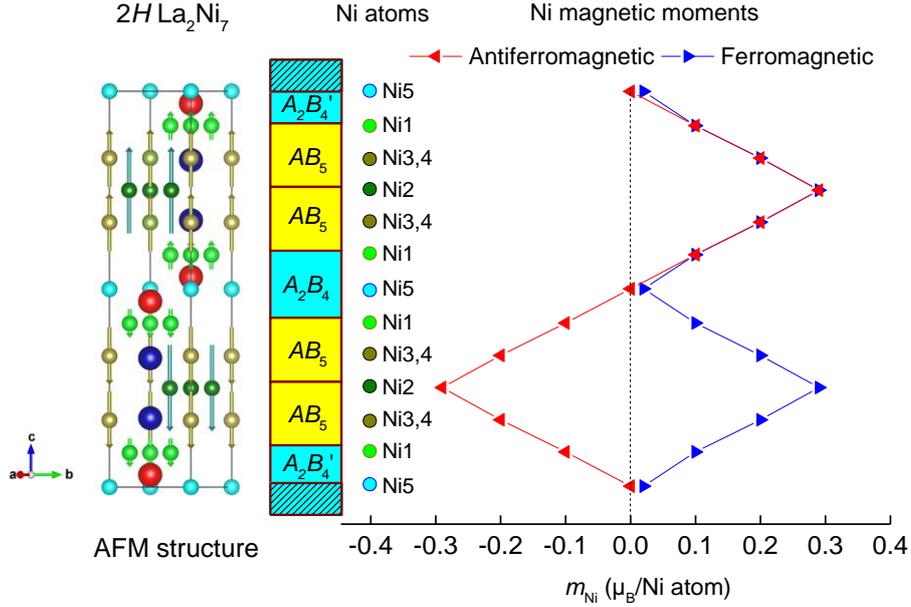

Figure. 7: Antiferromagnetic structure of La$_2$Ni$_7$ and Ni moment distribution along the *c* axis in both antiferromagnetic (red) and ferromagnetic (blue) structures of La$_2$Ni$_7$. (Color online).

The dependence of the Ni magnetic moments versus their localization in [$AB_2$] and [$AB_5$] units in hexagonal La$_2$Ni$_7$ can be partly related to the difference of magnetic properties of the corresponding binary LaNi$_2$ and LaNi$_5$ compounds: LaNi$_2$ is a Pauli paramagnet with a weak susceptibility ($\chi$ = 1.5 10$^{-4}$ emu/mole) [33] whereas LaNi$_5$ is an enhanced Pauli paramagnet with $\chi$ = 2.0 10$^{-3}$ emu/mole [50]. In general, $A$Ni$_5$ compounds are close to the onset of ferromagnetism, but the $N(E_F)$ is not large enough to stabilize a ferromagnetic state. The weak itinerant magnetism observed in $A_2$Ni$_7$ as well as $A$Ni$_3$ compounds has been attributed to the existence of the sharp and narrow peak near E$_F$ providing a large enough $N(E_F)$ contribution to stabilize a SP state [51].

However, all the experimental studies have shown that the magnetic ground state of 2*H*-La$_2$Ni$_7$ is characteristic of an AFM structure ($T_N$ = 50 K). In addition, 2*H*-La$_2$Ni$_7$ undergoes a metamagnetic transition towards a FM state with a transition field of 4.6 T and a stabilization of a wFM behavior above 6 T at 4.2 K [27]. Up to now, all the studies performed on La$_2$Ni$_7$ did not allow to propose an AFM structure filling these criteria. The only observation obtained from the DOS calculation was the smaller $N(E_F)$ value of La$_2$Ni$_7$ compared to Y$_2$Ni$_7$ [46].

Taking into account the symmetry of the hexagonal structure (2 blocks of [$AB_2$ + 2 $AB_5$]) and assuming that the Ni5 moments, which belong to the [$AB_2$] unit and are already very small in the calculated FM structure ($m_{Ni5}$ = 0.023 µB/Ni) are equal to zero, the AFM structure presented in figure 7 can be proposed . The spins of the Ni atoms are of same sign in the same block (between layers at $z$ = 0 and 1/2) whereas they have opposite sign in the next stacking (between $z$ = 1/2 and 1), and they are all aligned parallel to the *c* axis. In other words, Ni atoms in the same block are ferromagnetically aligned while the magnetic coupling between adjacent blocks is antiferromagnetic (Table 1, figure 7). The DFT calculation converges to the proposed AFM structure with a net magnetization equal to zero on the Ni5 site and an antiferromagnetic coupling between the Ni1 atoms which are its nearest neighbors and belonging to the different blocks ($z$ < ½ or $z$ > ½). Due to the particular point symmetry of the Ni5 site ($\bar{3}m.$), the molecular field generated by the six Ni1 neighbor atoms forming two tetrahedra with antiparallel spin cancels exactly at this 2*a* position (figure 5, left). This can explain that these Ni5 atoms have no ordered moments in the AFM structure. On the contrary in the FM configuration, where all the Ni moments are parallel to the *c* axis a small Ni moment is induced on the 2*a* position by the magnetic moments of the 6 Ni1 neighbors as explained above (figure 5, right).

Similar geometric environment with a central atom surrounded by 6 neighbors, which can be coupled ferromagnetically or antiferromagnetically, depending on the presence of an ordered moment or not on the central atom position, has been observed in other Laves phase systems: *C*14 hexagonal $TiFe_2$, $Hf_{0.825}Ta_{0.175}Fe_2$ and monoclinic $YFe_2H_{4.2}$ derived from a C15 cubic parent compound [52-54].

In $La_2Ni_7$, the distance between two next Ni1 neighbors of different blocks connected via a Ni5 atom is 5.20 Å and these distances are large enough to favor negative $\underline{J}_{Ni1-Ni1}$ interactions in the absence of a moment on Ni5 site. Except for Ni5 atoms, all the other Ni atoms carry Ni moments whose magnitudes are close to those calculated in the wFM state (Table 1, figure 4). As their environment remains similar in both FM and AFM structures, they are less affected by the magnetic ordering contrary to the Ni5 atoms.

The DOS for this AFM structure shows that the narrow peak observed in NSP DOS is shifted below $E_F$ for the majority spin and $N(E_F)$ falls in a valley at $E_F$ (figure 3c). However, only from the energy calculated for the proposed AFM and the FM structures, it is not possible to distinguish the most stable state since the difference is negligible with less of 1 meV/at., whereas the DFT accuracy is known to be few meV/at.. Thus, whatever the XC functional or method, the DFT calculation indicates that both magnetic structures present the same stability at 0 K. In Table 2, the total energy difference of $A_2Ni_7$ compounds is compared in NSP, FM and AFM structures for both the rhombohedral and hexagonal symmetries. It shows that the most stable state is the magnetic state (AFM or FM) for both $La_2Ni_7$ and $Y_2Ni_7$ without a clear difference between both symmetries, as already discussed in previous paper about stacking phases [33].

The DFT calculation allows to propose an hexagonal AFM structure, whereas it become more challenging to propose an AFM structure for the rhombohedral cell. In fact, the AFM ordering requires an even number of blocks to cancel the total magnetic moment, as observed in the $Ce_2Ni_7$ type structure. As the $Gd_2Co_7$ type structure contains three blocks, it will require a doubling of the unit cell along the *c* axis to obtain a comparable AFM order. As the *c* parameter of 3*R*-$Y_2Ni_7$ is already 36.19 Å, this will generate a very anisotropic magnetic cell with $c/a \approx$ 15, compared to only $c/a \approx$ 5 for 2*H*-$La_2Ni_7$. Although we have not calculated such large magnetic cell, it was observed experimentally that 3*R*-$La_2Ni_7$ should have a ferromagnetic ground state. Indeed, the magnetization curves of $La_2Ni_7$ sample annealed at 873 K and containing a mixture of hexagonal and rhombohedral phases, were analyzed by a superposition of FM and AFM states [27].

This proposed AFM structure for 2*H*-$La_2Ni_7$ allows explaining several experimental features of the literature as detailed below:

i) it confirms the high uniaxial anisotropy determined from the magnetization curves of 2*H*-$La_2Ni_7$ by Parker et al [27]. They calculated an anisotropy field of at least 12 T and an uniaxial anisotropy energy greater than 160 kJ/m$^3$.

ii) It can explain the positive value of $\theta_P$ above $T_N$ [26, 27, 30, 33], as locally the FM interactions remain larger than the AFM ones. Parker et al [27], already suggested that the large value of $\theta_P$ was characteristic of a metamagnetic behavior of a compound which has uniaxial anisotropy and with ferromagnetic intralayer and antiferromagnetic interlayer exchange interactions.

iii) The study of 2*H*-$La_2Ni_7$ by neutron powder diffraction (NPD) did not allow to observe magnetic lines at low temperature [33]. A simulation of the NPD pattern was therefore done for FM and AFM structures assuming a wavelength of 2.43 Å, typical of a spectrometer dedicated for magnetic structure determination (figure S2).

This reveals how it will be difficult to observe the magnetic structures experimentally. As the magnetic cell has a [0 0 0] propagation vector, it would only increase very weakly few nuclear line intensities. The most intense AFM magnetic line will contribute to only 1.2 % of the (1 0 0) Bragg peak and even less considering the experimental moments and not the calculated one.

Although the DFT calculations cannot observe a clear difference of stability between the AFM and FM structures, all the experimental magnetic studies clearly indicates that the ground state of 2$H$-La$_2$Ni$_7$ is AFM. But the difference of stability between the two types of magnetic order is clearly small: it was observed experimentally that the substitution of only 3 at % of Cu for Ni stabilizes a FM state, whereas the AFM ground state is maintained up to 10 at% Co [29]. This difference was attributed to a critical number of $d$ electrons as the total number of $d$-electron increases upon Cu substitution and decreases upon Co substitution. It confirms that we are very close to the limit of stability of the AFM structure, which can be favored by the existence of spin-fluctuations. The low value of the saturation moment and the large Rhodes-Wohlfarth ratio observed for both 2$H$-La$_2$Ni$_7$ [27, 29, 30, 33] and 3$R$-Y$_2$Ni$_7$ compounds [25] as well as other experimental results confirms their weak itinerant character and the existence of large spin fluctuations in both compounds. In addition, the metamagnetic behavior of the magnetization curves of La$_2$Ni$_7$ was analyzed within the Moriya and Usami theory using a Landau-type free energy equation, used in the frame of systems with large spin fluctuations [30, 55, 56].

For weak itinerant magnetic systems, the existence of large spin fluctuations was able to explain the difference between results obtained by DFT calculations and those observed experimentally. For example, Ni$_3$Ga is a strongly renormalized paramagnet, but the DFT calculation predicts a weak ferromagnetic ground state similar to isostructural Ni$_3$Al, with even larger Ni moment [9]. This difference has been explained by taking into account the low-frequency spin fluctuations present in this compound which is close to a ferromagnetic quantum critical point. The theoretical study of TiAu confirmed also the influence of spin fluctuations for the stabilization of the weak antiferromagnetic state, as well as the overestimation of the Ti moment determined by DFT calculations [19]. Such overestimation of magnetic moment is often observed for compounds which lies close to quantum critical point.

In the case of 2$H$-La$_2$Ni$_7$ the magnetic order is driven by the magnitude of the Ni5 moment which is predicted by DFT calculation to be 0.023 μ$_B$ in the FM structure and 0 in the AFM structure. But it was found, that the total magnetic moment of 2$H$-La$_2$Ni$_7$ in FM state is overestimated of at least 30 % compared to the experimental value. In this case, the influence of spin fluctuations on the particular Ni5 site should be large enough to stabilize a geometric environment with the next Ni1 neighbors adopting an antiparallel coupling.

## 4. Conclusions

In this work, we have proposed for hexagonal La$_2$Ni$_7$ an AFM structure with a stability comparable to that of the FM structure. This AFM structure of 2$H$-La$_2$Ni$_7$ results from the geometric conditions presented by its stacked hexagonal structure: the molecular field is canceled on the Ni5-2$a$ site at $z = 0$ and $z = 1/2$ which belong to the [$AB_2$] units and acts as in inversion center for the first neighbors Ni at the interface between the [$AB_2$] and [$AB_5$] units (Ni1-12$k$). These nearest Ni atoms adopt an AFM coupling and the symmetry of the structure on this position allows an inversion of the sign of the other Ni spin orientation belonging to different blocks. This generates a modulated AFM structure with two ferromagnetic slabs of opposite directions separated by a non-magnetic layer. Such magnetic structure type was already observed in some $A$Fe$_2$ Laves phases or their hydrides. The metamagnetic transition from AFM towards a FM structure is explained by a non-zero Ni moment on the Ni5 site induced by the application of a magnetic field.

The stabilization of this AFM structure, observed experimentally, should be explained by the existence of large spin fluctuations, not taken into account by simple DFT calculation, as observed for other weak itinerant magnets as Ni$_3$Ga or TiAu. Further works using Moriya and Usami theory of spin fluctuation, could be very interesting to consider for La$_2$Ni$_7$.

Further experimental and theoretical studies will be performed on pseudo-binary La$_{2-x}$Y$_x$Ni$_7$ compounds to observe the evolution from wAFM 2$H$-La$_2$Ni$_7$ towards wFM 3$R$-Y$_2$Ni$_7$, and verify what the critical geometric and electronic parameters are. First principles calculations of other compounds with stacking structures like $A$Ni$_3$ or $A_5$Ni$_{19}$ compounds from [$A_2$Ni$_4$ + $n.A$Ni$_5$] with $n = 1$ and 3 respectively, could be also very interesting in order to analyze the relation between their crystal structures and magnetic properties.

**Acknowledgements**

The authors are very thankful to M. Gupta for her advice and fruitful discussion. DFT calculations were performed using HPC resources from GENCI-CINES (Grant 2018-96175).